\documentstyle[aps,prl,twocolumn,epsf,floats]{revtex}

\begin{document}

\draft

\title{Quantum Topological Excitations:
from the Sawtooth Lattice to the Heisenberg Chain}

\author{S. A. Blundell$^{1}$ and M. D. N\'{u}\~{n}ez-Regueiro$^{2}$}

\address{%
$^{1}$D\'epartement de Recherche Fondamentale sur la Mati\`ere
Condens\'ee, CEA-Grenoble/DSM \\
17 rue des Martyrs, F-38054 Grenoble Cedex 9, France\\
$^{2}$Grenoble High Magnetic Field Laboratory,
MPI-FKF and CNRS, B.P.\ 166, 38042 Grenoble, Cedex 9, France}

\date{\today}

\maketitle

\begin{abstract}
The recently elucidated structure of the delafossite YCuO$_{2.5}$
reveals a Cu-O network with nearly independent $\Delta$ chains having
different interactions between the $s=1/2$ spins.  Motivated by this
result, we study the $\Delta$ chain for various ratios $J_{\rm
bb}/J_{\rm bv}$ of the base-base and base-vertex interactions.  By
exact diagonalization and extrapolation, we show that the elementary
excitation spectrum, which (within numerical error) is the same for
total spins $S_{\rm tot}=0$ and 1, has a gap only in the interval
$0.4874(1) \leq J_{\rm bb}/J_{\rm bv} \leq 1.53(1)$.  The gap is
dispersionless for $J_{\rm bb}/J_{\rm bv}=1$, but has increasing
$k$-dependence as $J_{\rm bb}/J_{\rm bv}$ moves away from unity,
related to the instability of dimers in the ground state.
\end{abstract}


There is currently great interest in the topological excitations of
quantum antiferromagnetic (AF) spin systems.  The isotropic $s=1/2$
Heisenberg chain has spinon excitations, while the anisotropic chain
in its ordered phase has localized domain walls between opposite AF regions,
both widely discussed in the literature.  Later on, Shastry and Sutherland
\cite{shastry} introduced another class of excitations: topological
quantum solitons separating different regions of broken translational
symmetry.  The typical example is the symmetric
zigzag spin ladder, first addressed by Majumdar and Ghosh
\cite{majumdar} (MG), in which nearest-neighbor (NN) triangles sharing
a base site are also vertex-vertex coupled.  Its lowest-energy
excitations are so-called {\em kinks} (K) and {\em antikinks} (AK),
both with similar characteristics and giving rise to a finite gap
$\Delta E
\approx 0.234J_{1}$ when the interaction $J_{2}$ between
next-nearest-neighbors (NNN) is half that of NN spins $J_{1}$
\cite{caspers}.

Then attention turned to the sawtooth or $\Delta$ chain, which
consists of coupled $s=1/2$ Heisenberg spins forming triangles aligned
in a chain with a common base site, like the zigzag ladder but without
the vertex-vertex coupling.  The $\Delta$ chain has unique properties.
In fact, numerical \cite{nakamura} and variational
\cite{sen} studies of this lattice, all with bonds having the same
interaction, have shown a remarkable feature: the K-AK symmetry of the
MG model is broken here, yielding however a similar dispersionless
small gap for the low-lying excitation modes.  Very recently, in a
study of an asymmetric zigzag ladder with NNN alternating
interactions, that is, with $J_{2}\pm\delta$, Chen {\em et al.}
\cite{chen} have discussed the crossover from the MG model
($J_{2}=J_{1}/2$, $\delta=0$) to the $\Delta$ chain with equal
interactions ($\delta=J_{2}=J_{1}/2$).

Theoretical interest in the sawtooth lattice itself is now enhanced by
new experimental results.  In fact, the overdoped $R$CuO$_{2+x}$ ($R=$
Y, La, etc.)  delafossite compounds, first synthesized by Cava {\em et
al.}\ \cite{cava}, opened up new possibilities for studying hexagonal
Cu planes with AF interactions between the Cu$^{2+}$ ions.  Depending
on the O-doping, different $s=1/2$ effective magnetic lattices are
obtained, although with much weaker interactions than high-$T_{c}$
systems, which have comparable bond lengths but $180^{\circ}$ Cu-O-Cu
angles.  Studies \cite{simon,nunez} of the diluted kagom\'e lattices
of the $x=0.66$ case predicted interesting properties.  The detailed
structure of YCuO$_{2.5}$ has just been elucidated \cite{tendeloo,garlea}
and, as suggested earlier
\cite{nakamura,sen}, this compound appears as a physical realization of
the sawtooth lattice (Fig.\ \ref{fig1}).  Recent synthesis
\cite{garlea,tendeloo} has succeeded in obtaining an orthorhombic 2H single
phase, in which the additional $x=0.5$ oxygen ions are located at the
center of alternating sets of triangles, providing super-exchange
paths between $s=1/2$ spins on nearly independent $\Delta$ chains.
However, careful consideration of the measured angles and distances
indicates different interactions between two spins on the base $J_{\rm
bb}$ and between the base-vertex NN spins $J_{\rm bv}$ of the
triangles, most probably with $J_{\rm bb}<J_{\rm bv}$, owing to the
smaller Cu-O-Cu angle
\cite{tendeloo}.  While the case $J_{\rm bb}=J_{\rm bv}$ has been quite
extensively studied theoretically \cite{nakamura,sen}, to the best of
our knowledge the case $J_{\rm bb}\neq J_{\rm bv}$ has never been considered
before.

Therefore, we here analyze the sawtooth lattice for various ratios
J$_{\rm bb}$/J$_{\rm bv}$ of these AF couplings.  The Hamiltonian is
given by
\begin{eqnarray}
H&=&J_{\rm bb}\sum_{i=1}^{N}{\bf s}_{2i-1}\cdot{\bf s}_{2i+1}+
\nonumber\\
    && J_{\rm bv}\sum_{i=1}^{N}({\bf s}_{2i-1}\cdot{\bf s}_{2i}+
                 {\bf s}_{2i}\cdot{\bf s}_{2i+1}),
\label{eqn:hamiltonian}
\end{eqnarray}
where $N$ is the number of triangles ($2N$ spins) in the chain, and
${\bf s}_{i}$ is the spin-$1/2$ operator at site $i$. Note that when
$J_{\rm bb}=0$ we retrieve the
Heisenberg chain with $2N$ sites, while when $J_{\rm bv}=0$ we have a
Heisenberg chain with $N$ sites and $N$ disconnected spins.  It is
therefore interesting to study the evolution of the elementary
excitations from the sawtooth lattice with equal interactions $J_{\rm
bb}=J_{\rm bv}$, in which K-AK pairs yield a dispersionless small gap
\cite{nakamura,sen} of the same order as that in the symmetric zigzag
ladder, to the isotropic Heisenberg chain, in which pairs of spinons
exhibit a strongly dispersive spectrum \cite{dCP} with excitation
energy varying between zero and $\pi J/2$ ($J$ is the coupling
between adjacent spins).

To test our numerical procedure, which includes larger clusters than
before (up to 12 triangles), and to make contact with previous theory,
we first briefly reconsider, and extend, existing numerical results for
the $\Delta$ chain with $J_{\rm bb}=J_{\rm bv}=J$.
In this case, Eq.\ (\ref{eqn:hamiltonian}) has two degenerate
ground states with $N$ dimers \cite{monti}.  These ground states may
be written as states in which each spin on the base of a triangle
forms a singlet either with the following vertex spin (right, R-dimer
state) or with the previous vertex spin (left, L-dimer state), that
is,
\begin{eqnarray}
|{\rm GS, R} \rangle &=& \prod_{i=1}^{N}[2i-1,2i], \label{eqn:Rdimer}\\
|{\rm GS, L}\rangle &=& \prod_{i=1}^{N}[2i,2i+1], \label{eqn:Ldimer}
\end{eqnarray}
where $[i,
j]\equiv(|\alpha_{i}\beta_{j}\rangle-|\beta_{i}\alpha_{j}\rangle)/\sqrt{2}$,
with $\alpha_{i}$ ($\beta_{i}$) denoting the states with
$s_{i}^{z}=1/2$ ($-1/2$) at the site $i$.  These two states are
linearly independent and become orthogonal for $N\rightarrow\infty$.
The existence of an excitation gap was also rigorously proved
\cite{monti}.  The elementary excitations have been shown
\cite{nakamura,sen} to be well-separated K-AK-type domain walls
separating regions of R-dimers and L-dimers.  A K has a dimer in its
triangle, while an AK does not.  Curiously, they have very different
characteristics in this system.  A K has no excitation energy and is
localized, but an AK propagates with kinetic energy within a region
bounded by kinks.  As a consequence of the former property the
low-lying excitation spectrum is dispersionless, and owing to the
second one, the gap value is considerably reduced compared to the
energy of a trivial triplet state replacing a singlet dimer of the
ground state.

We diagonalize the spin Hamiltonian $H$ in Eq.\
(\ref{eqn:hamiltonian}) by the Lanczos algorithm, using periodic
boundary conditions with ${\bf s}_{2N+1}$ identified with
${\bf s}_{1}$.  All sizes from $N=4$--12 triangles are calculated.
The values of the excitation energies obtained are then extrapolated
to the $N\rightarrow\infty$ limit by assuming the
finite-size error term to be a polynomial in $1/N$, whose coefficients
are determined by fitting.  In Table \ref{tab1} we give the gaps found
for wavevectors $k=0$, $\pi/2$, and $\pi$ for excitations with total
spin $S_{\rm tot}=0$, 1, and 2, for the case $J_{\rm bb}=J_{\rm
bv}=J$.  (Note that the wavevector $k$ of an eigenstate $|\psi\rangle$
is here defined such that $T_{n} |\psi\rangle = e^{ikn}|\psi\rangle$,
where $T_{n}$ is the translation operator by $n$ triangles or $2n$
spins.)  The quoted numerical error, which arises entirely from the
extrapolation, is defined (here and later) to be twice the change in
the result upon discarding the data of the largest system calculated
(12 triangles) and repeating the extrapolation.  In this case, the
leading finite-size effect goes as $1/N^{2}$.  Table \ref{tab1}
confirms the gap to be dispersionless, within numerical error.  Our
most precise value of the gap, for $k=0$ and $S_{\rm tot}=1$, is
$\Delta E = 0.2155(3) J$, confirming the estimate 0.215 by Nakamura
and Kubo
\cite{nakamura}, and a little smaller than the value 0.219 expected
from the variational calculation by Sen {\em et al.}\ \cite{sen} for an
AK configuration spread over 5 sites.  We find that, within numerical
error, the $S_{\rm tot}=0$ low-lying excitations become degenerate with
the $S_{\rm tot}=1$ gap as $N\rightarrow\infty$. Furthermore, the
spectrum for $S_{\rm tot}=2$ appears also to be dispersionless,
but about twice that for $S_{\rm tot}=1$ or $0$,
as shown in Table \ref{tab1}.  This new result is
contrary to the speculation in Ref.\ \cite{kubo} that the
excitation energies for higher spins might converge
to the same limiting value as $N\rightarrow\infty$.

Figure 2 shows the evolution of the low-lying triplet ($S_{\rm
tot}=1$) excitation spectrum for $J_{\rm bb}/J_{\rm bv}\leq 1$.  As
$J_{\rm bb}$ decreases, the triplet excitation energy decreases at
$k=0$ until it vanishes near $J_{\rm bb}/J_{\rm bv}\approx 0.5$, while
for $k=\pi$ it goes up.  Progressively a stronger $k$-dispersion
appears, yielding for $J_{\rm bb}=0$ the famous lower boundary
expression for the continuum of excited triplet states for the
isotropic $s=1/2$ Heisenberg chain, $\Delta E_{L}(k)= (\pi/2) J_{\rm
bv}|\sin k/2|$ \cite{dCP} (here rewritten keeping the definition of
$k$ for our $\Delta$ chain).

For $J_{\rm bb}=J_{\rm bv}$ the ground state is doubly degenerate for
all $N$.  But for $J_{\rm bb}<J_{\rm bv}$ and $N$ finite, these states
are split into a ground state with spin-parity $0^{+}$ and a first
excited singlet state with spin-parity $0^{-}$.  However, we find that
the latter approaches the ground state with asymptotic dependence
$1/N$ as as $N\rightarrow\infty$, restoring the double degeneracy.

An accurate determination of the critical ratio $J_{\rm bb}/J_{\rm
bv}<1$ required to produce a triplet gap may be made by the method
proposed in Ref.\ \cite{black}.  The idea is to map this problem to a
continuum field theory and take into account the fact that in a
fermion system with a fixed number of particles umklapp scattering is
the only interaction that splits the degeneracy of the two lowest
excited states.  Therefore, the difference of their energies provides
a precise measure of the umklapp processes, which vanish at the
critical interaction ratio.  For the $\Delta$ chain, with $N$ finite
and $J_{\rm bb}<J_{\rm bv}$, the first two excited states have
spin-parity $0^{-}$ and $1^{+}$ and undergo a level-crossing near
$J_{\rm bb}/J_{\rm bv}\approx 0.5$.  As shown in Fig.\ \ref{fig3}, the
value of $J_{\rm bb}/J_{\rm bv}$ at the level crossing follows a
polynomial in $1/N^{2}$, and its extrapolation to $N\rightarrow\infty$
yields $(J_{\rm bb}/J_{\rm bv})_{\rm crit}=0.4874(1)$.

The ground state when $J_{\rm bb}=J_{\rm bv}$ consists of spin
singlets between NN spins, the R-dimers or L-dimers of Eqs.\
(\ref{eqn:Rdimer}) and (\ref{eqn:Ldimer}).  To study the evolution of
the ground state $|0\rangle$ for $J_{\rm bb}<J_{\rm bv}$, we plot in
Fig.\ \ref{fig4} the {\em dimerization fraction}, defined as $D_{\rm
frac}=|\langle 0|{\rm GS,R}\rangle|^{2}+|\langle 0|{\rm
GS,L}\rangle|^{2}$
\cite{dimerfrac}.  $D_{\rm frac}$ drops from its value of unity
at $J_{\rm bb}=J_{\rm bv}$ toward zero as $J_{\rm bb}/J_{\rm bv}$
decreases, showing that the gap reduction is related to
increasing fluctuations of the dimer state.  Fig.\ \ref{fig4} shows
how important the extrapolation to $N\rightarrow\infty$ is in this
calculation; we found that an Aitkens extrapolation gave very stable
results.

Turning to the case $J_{\rm bb}/J_{\rm bv}>1$, we show in
Fig.\ \ref{fig5} how the $S_{\rm tot}=1$ gap dispersion curves are modified.
The minimum gap is now found for $k=\pi$ and decreases with increasing
interaction ratio.  On the other hand, for large interaction ratios
the low-lying states become nearly degenerate.  This is easy to
understand: we are again approaching the Heisenberg chain, though just
for the $N$ sites on the bases of the triangles, while the remaining
$N$ spins on the vertex are only loosely coupled, leading to a complex
of $2^{N}$ nearly degenerate states.  The evaluation of the critical
ratio for the closure of the gap also becomes more complicated
(details will be given elsewhere), but an approach related to that
used above gives $(J_{\rm bb}/J_{\rm bv})_{\rm crit}=1.53(1)$.

Figure 6 summarizes our main numerical results for the low-lying
excitations of the sawtooth lattice.  A finite gap is found
only for interaction ratios within the interval $0.4874(1) \leq J_{\rm
bb}/J_{\rm bv} \leq 1.53(1)$.  Thus, the curve is not completely
symmetric around $J_{\rm bb}/J_{\rm bv}=1$.
While our above discussion of the excitation gap has been explicitly
just for $S_{\rm tot}=1$ excitations, we note that we have also been
able to calculate many features of the spectra for $S_{\rm tot}=0$
with a numerical accuracy of better than a few percent, finding
agreement with the $S_{\rm tot}=1$ values in all cases.  Examples: the
excitation energy at $k = \pi$ for $0 \leq J_{\rm bb}/J_{\rm bv}
\lesssim 1.5$ and at $k = 0$ for $0.9 \leq J_{\rm bb}/J_{\rm bv} \leq
1.0$.  Also, for $S_{\rm tot}=0$ gap closure we find $(J_{\rm
bb}/J_{\rm bv})_{\rm crit} = 1.51(3)$ compared to $(J_{\rm bb}/J_{\rm
bv})_{\rm crit} = 1.53(1)$ for $S_{\rm tot}=1$.  Further details will
be published elsewhere.  This provides strong numerical evidence that
the lowest excitation spectra are in fact four-fold degenerate (in the
limit $N \rightarrow
\infty$) for all $0
\leq J_{\rm bb}/J_{\rm bv} \lesssim 1.5$, thus generalizing the known
result for the isotropic $s = 1/2$ Heisenberg chain ($J_{\rm bb} = 0$)
\cite{fazekas}.

To conclude, we have shown that different interactions along the base
and vertex bonds in the sawtooth lattice yield a $k$-dependent
elementary excitation spectrum, the effective gap being progressively
reduced with increasing instability of the dimer ground state.  The
recent crystallographic study of YCuO$_{2.5}$ gives values for the
three sides of the triangles and for the Cu-O-Cu angles.  The
consideration of these parameters and of the different coordination of
the Cu ions (see Fig.\ \ref{fig1} and Ref.\ \cite{tendeloo}) justifies
taking the same base-vertex interaction $J_{\rm bv}$ for both sides of
the triangle, but a different $J_{\rm bb}$ along the base, as done
here.  The effect of a stronger AF interaction in just one of the
vertex bonds is quite clear: it will lift the degeneracy of the gapped
ground state, stabilizing the dimer singlet ordering corresponding to
that direction.  Preliminary susceptibility measurements on YCuO$_{2.5}$
by Garlea and co-workers \cite{garlea} yield a low-temperature gap
$\sim17$~K, which would imply AF interactions of $\sim$80~K if one
assumes $J_{\rm bb}=J_{\rm bv}$, a very reasonable value for this
system. Now that good samples are available, measurements using
more appropriate methods are in progress to distinguish the different
interactions and to obtain a precise value of the gap, if one exists
(depending on the interaction ratio).  It will be worthwhile to
synthesize single crystals to allow the study of the dispersion of the
excitation spectrum.

\begin{figure}[tb]
\caption{Sawtooth chains in the triangular Cu planes of the
delafossite YCuO$_{2.5}$.  The extra O ions (white) for $x=0.5$ are
located at the center of particular triangles of Cu ions, creating AF
super-exchange only within these triangles.  This gives nearly
independent $\Delta$ chains, indicated by thin black lines.  While Cu1
(black) adopts tetrahedral coordination with two O ions in this plane
(and two other O out of it), Cu2 (gray circles) adopts triangular
coordination with just one O in this plane.  The angles and distances
\protect\cite{tendeloo} suggest a weaker $J_{\rm bb}$ interaction
between Cu1-O-Cu1 bonds (bases of the triangles) than for the
Cu1-O-Cu2 base-vertex bond $J_{\rm bv}$.}
\label{fig1}
\end{figure}

\begin{figure}[b!]
\caption{Low-lying excitation spectra of the sawtooth lattice for
$J_{\rm bb}/J_{\rm bv}<1$.  The dashed line corresponds to the $J_{\rm
bb}=J_{\rm bv}$ case, while the $J_{\rm bb}=0$ dotted-dashed curve
follows the dispersion of the isotropic Heisenberg chain
(using here the definition of $k$ for the $\Delta$ chain).}
\label{fig2}
\end{figure}

\begin{figure}[tb]
\caption{Variation of the critical interaction ratio for gap
closure when $J_{\rm bb}/J_{\rm bv}<1$, as a function of the number of
triangles $N$.  The polynomial extrapolation in $1/N^{2}$ yields the
best estimation of this critical value.}
\label{fig3}
\end{figure}

\begin{figure}[tb]
\caption{Dimerization fraction (see text) vs.\ the interaction ratio.
The calculations have been done for up to 12 triangles. The
importance of the extrapolation is shown.}
\label{fig4}
\end{figure}

\begin{figure}[b!]
\caption{Dispersion curve for the gap to $S_{\rm tot}=1$ excited
states of the sawtooth lattice with $J_{\rm bb}/J_{\rm bv}>1$,
compared to the $J_{\rm bb}/J_{\rm bv}=1$ case.}
\label{fig5}
\end{figure}

\begin{figure}[tb]
\caption{Gap to the lowest $S_{\rm tot}=1$ excited states of the
sawtooth lattice vs.\ $J_{\rm bb}/J_{\rm bv}$, after extrapolation to
$N\rightarrow\infty$.  The error is greater for $J_{\rm bb}/J_{\rm
bv}>1$ because states with $k=\pi$ exist only for even $N$, giving
less points in the extrapolation.}
\label{fig6}
\end{figure}

\begin{table}[tb]
\caption{Values of the
lowest-energy excitations of the sawtooth lattice for $J_{\rm
bb}=J_{\rm bv}$ with total spin $S_{\rm tot}$ and wavevector $k$,
after extrapolation to $N\rightarrow\infty$.}
\begin{tabular}{lddd}
$S_{\rm tot}$&gap ($k=0$)&gap ($k=\pi/2$)&gap ($k=\pi$)\\
\tableline
0&0.2153(8) &0.22(1)&0.216(2)\\
1&0.2155(3)&0.214(10)&0.216(2)\\
2&0.46(1)  &0.49(8) &0.46(6)\\
\end{tabular}
\label{tab1}
\end{table}


\begin{references}

\bibitem[*]{byline} E-mail: sblundell@cea.fr (SAB),
nunezreg@polycnrs-gre.fr (MDNR)

\bibitem{shastry} B. S. Shastry and B. Sutherland, Phys.\ Rev.\ Lett
{\bf 47}, 964 (1981).

\bibitem{majumdar} C. K. Majumdar and D. K. Ghosh, J.\ Math.\ Phys.\
{\bf 10}, 1388 and 1399
(1969); C. K. Majumdar, J.\ Phys.\ C {\bf 3}, 911 (1970).

\bibitem{caspers} W. J. Caspers, K. M. Emmett, and W. Magnus, J.\ Phys.\ A
{\bf 17}, 2687 (1984).

\bibitem{nakamura} T. Nakamura and K. Kubo, Phys.\ Rev.\ B {\bf 53}, 6393
(1996), and references therein.

\bibitem{sen} D. Sen, B. S. Shastry, R. E. Walsted, and R.J. Cava,
Phys.\ Rev.\ B {\bf 53}, 6401 (1996).

\bibitem{chen} S. Chen, H. Buttner and J. Voit, Phys.\ Rev.\ Lett {\bf 87},
087205 (2001); cond-mat/0201004.

\bibitem{cava} R. J. Cava {\em et al.},
J. Solid State Chem. {\bf 104}, 437 (1993).


\bibitem{simon} M. E. Sim\'{o}n, A. A. Aligia, and
M. D. N\'{u}\~{n}ez-Regueiro, Phys.\ Rev.\ B {\bf 51}, R15642 (1995).

\bibitem{nunez} M. D. N\'{u}\~{n}ez-Regueiro, C. Lacroix, and B. Canals,
Phys.\ Rev.\ B {\bf 54}, R736 (1996).

\bibitem{tendeloo} G. Van Tendeloo, O. Garlea, C. Darie, C. Bougerol-Chaillout,
and P. Bordet, J. Solid State Chem. {\bf 156}, 428 (2001).

\bibitem{garlea}O. Garlea, Thesis Grenoble University (2001); A.
Sulpice et al. (to be published).

\bibitem{dCP} J. des Cloizeaux and J. J. Pearson,
Phys.\ Rev.\ {\bf 128}, 2131 (1962).

\bibitem{monti} F. Monti and A. S\"{u}to, Phys.\
Lett.\ A {\bf 156}, 197 (1991); Helv. Phys. Acta {\bf 65}, 560 (1992).

\bibitem{kubo} K. Kubo, Phys.\ Rev.\ B {\bf 48}, 10552 (1993).

\bibitem{black} J. L. Black and V. J. Emery, Phys.\ Rev.\ B
{\bf 23}, 429 (1981); V. J. Emery and C. Noguera, Phys.\ Rev.\ Lett {\bf 60},
631 (1988).

\bibitem{dimerfrac} Since $|{\rm GS,R}\rangle$ and $|{\rm GS,L}\rangle$
are not orthogonal for $N$ finite (but are linearly independent),
   we should define $D_{\rm frac}=|\langle
0|R'\rangle|^{2}+|\langle 0|L'\rangle|^{2}$, where $|R'\rangle$ and
$|L'\rangle$ are 2 orthonormal states in the subspace spanned by
$|{\rm GS,R}\rangle$ and $|{\rm GS,L}\rangle$.

\bibitem{fazekas} P. Fazekas and P. S\"{u}to, Solid State Commun.
{\bf 19}, 1045 (1976).

\end{references}
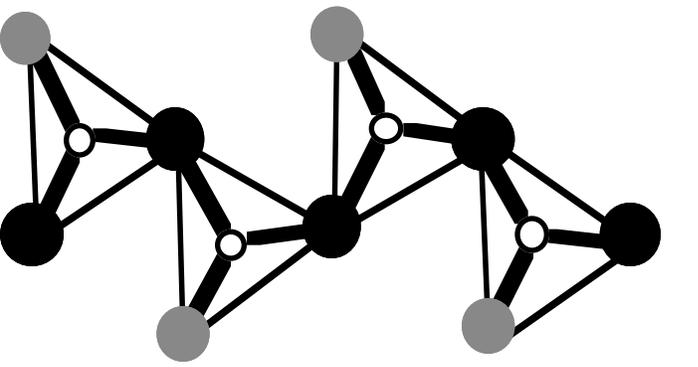
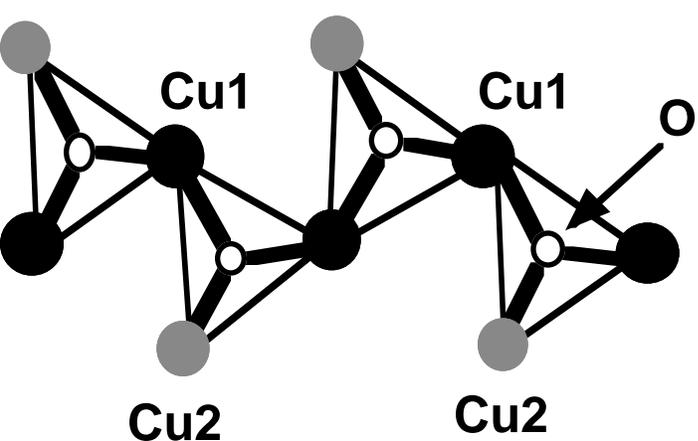
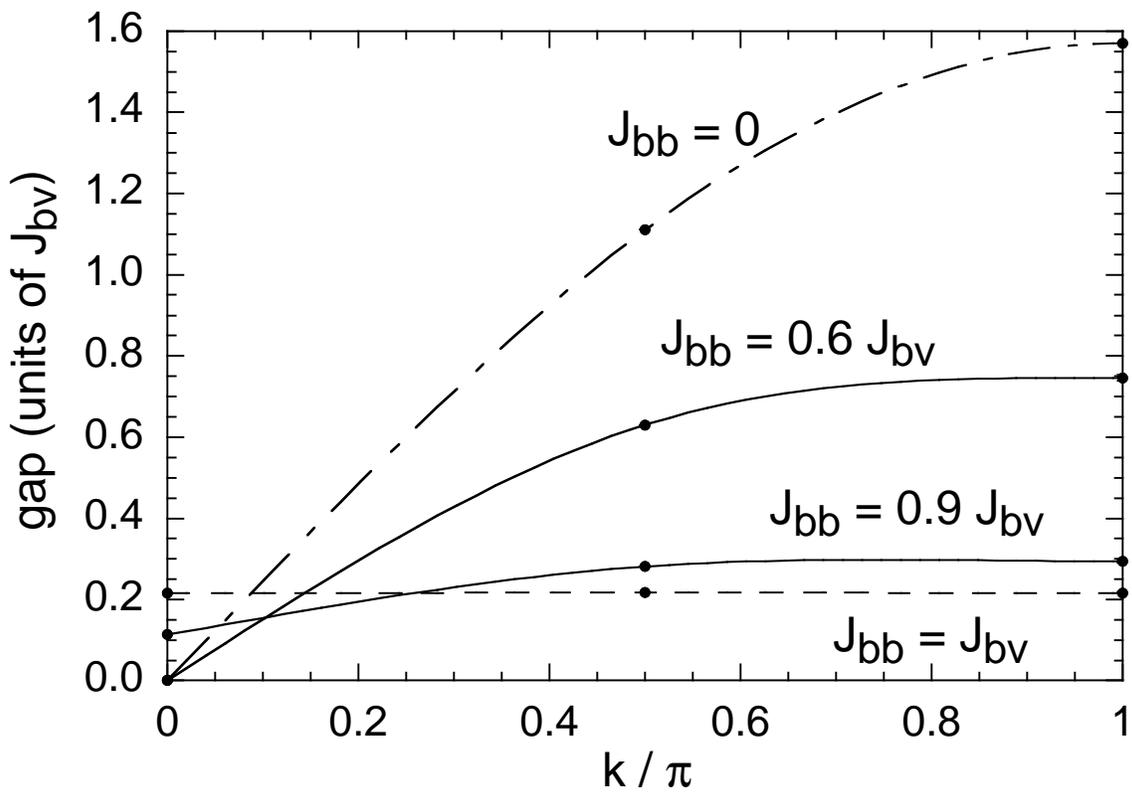
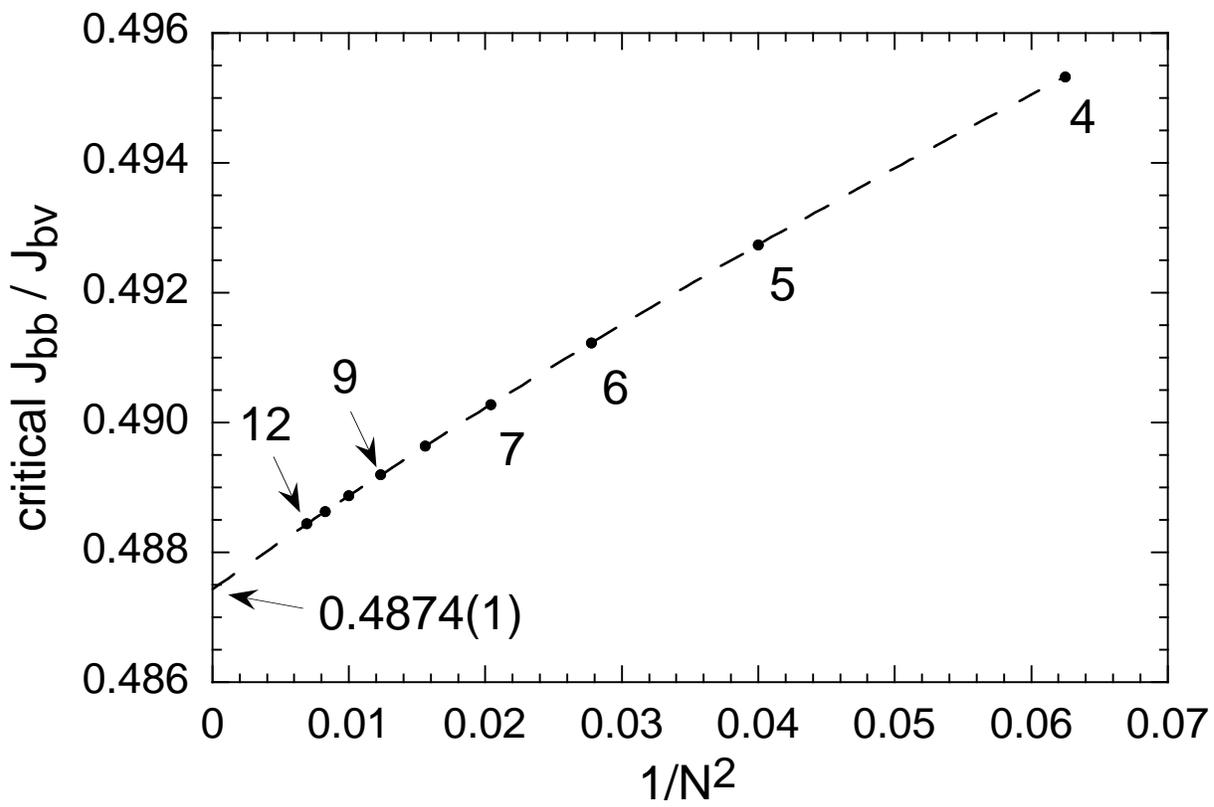
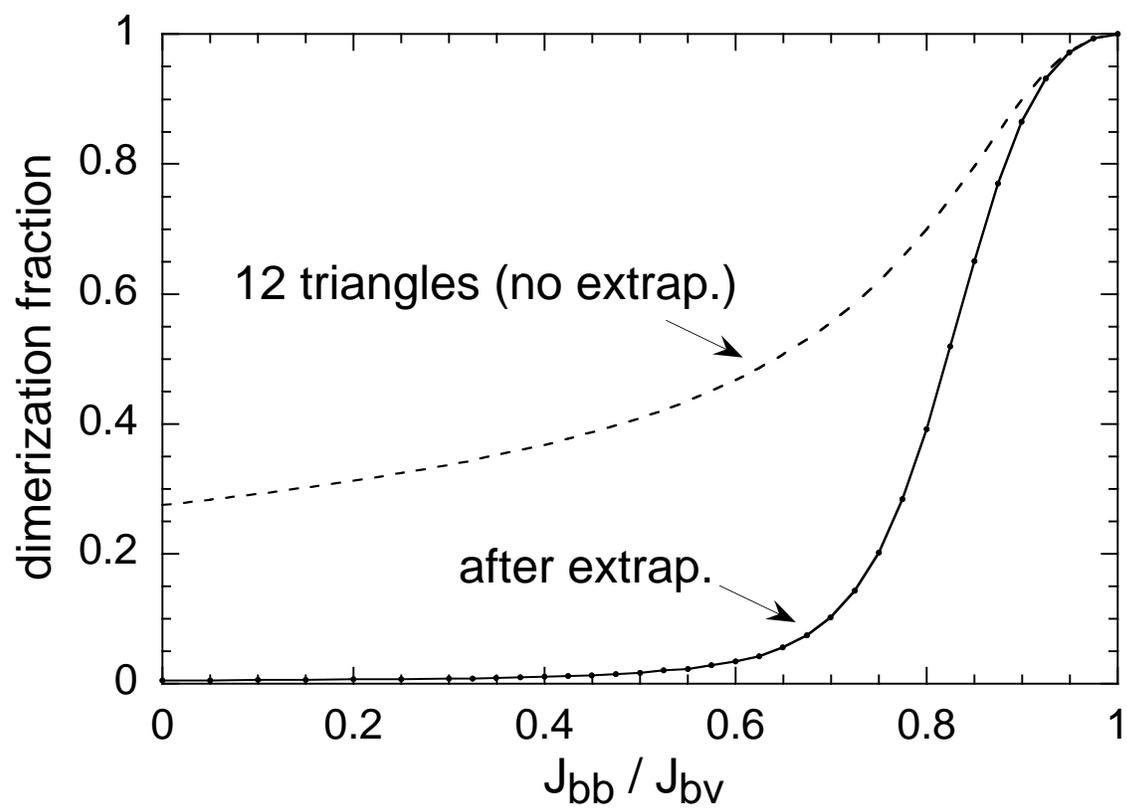
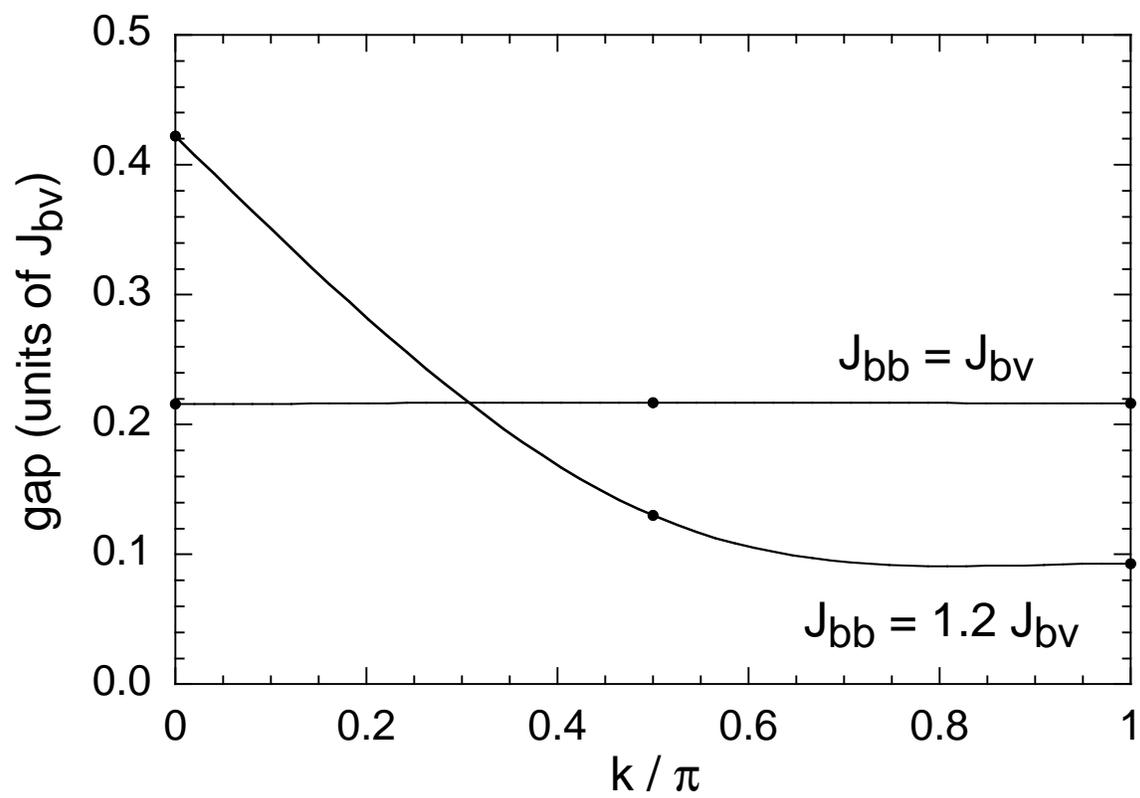
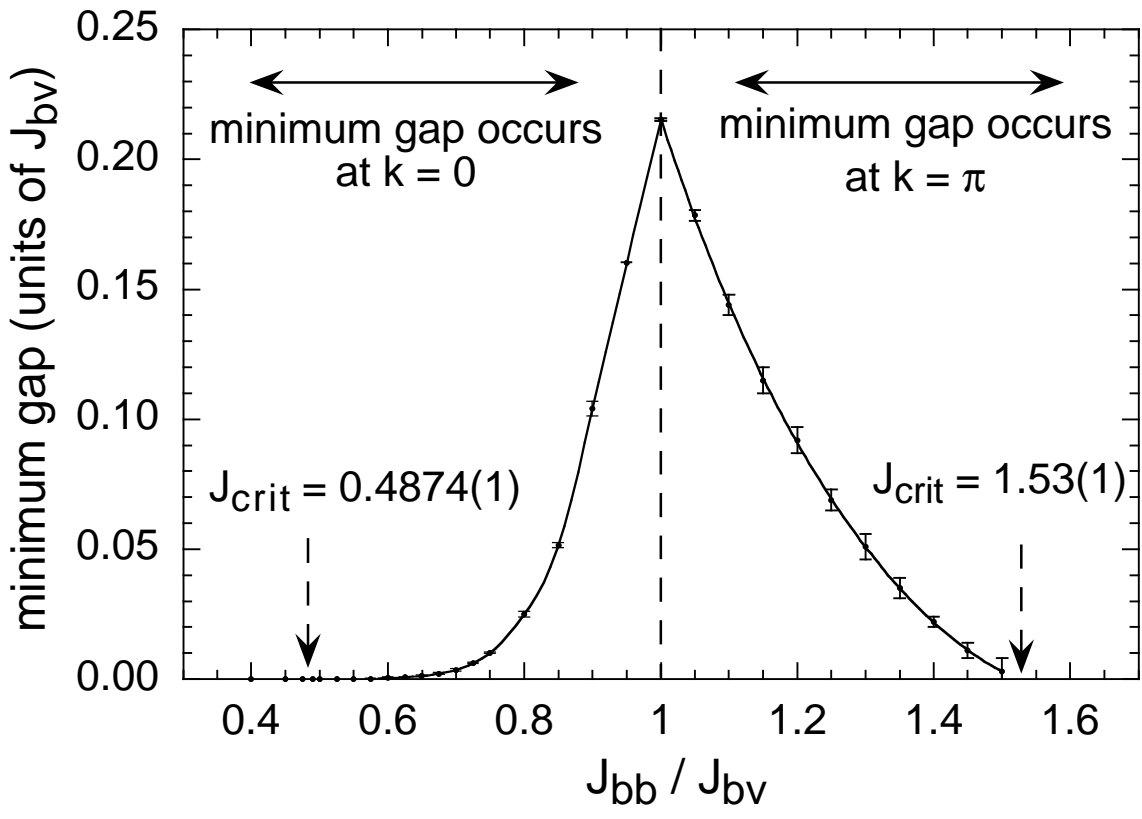
\end{document}